# On optimal allocation of treatment/condition variance in principal component analysis

André Beauducel & Norbert Hilger

Institute of Psychology, University of Bonn, Germany

19th April 2018

**Abstract**

The allocation of a (treatment) condition-effect on the wrong principal component (misallocation of variance) in principal component analysis (PCA) has been addressed in research on event-related potentials of the electroencephalogram. However, the correct allocation of condition-effects on PCA components might be relevant in several domains of research. The present paper investigates whether different loading patterns at each condition-level are a basis for an optimal allocation of between-condition variance on principal components. It turns out that a similar loading shape at each condition-level is a necessary condition for an optimal allocation of between-condition variance, whereas a similar loading magnitude is not necessary.

Institute of Psychology, University of Bonn, Kaiser-Karl-Ring 9, 53111 Bonn, Germany

Email: beauducel@uni-bonn.de; nhilger@uni-bonn.de

## 1. Introduction

*1.1 Condition effects in Principal Components Analysis*

Principal components analysis (PCA) has regularly been performed for the analysis of event-related potentials of the electroencephalogram (Dien, Khoe & Mangun, 2007; Dien, 2010; Kayser & Tenke, 2003, 2005). In the context of event-related potentials, PCA is often performed for observed variables representing *k* levels of at least one (experimental) condition factor, so that the components represent a mixture of the between- and within-condition variance. However, (experimental) condition factors occur in several areas of research and PCA is performed in several areas of research. It is therefore interesting to know how experimental condition effects are optimally allocated on principal components.

*1.2 Misallocation of between-condition variance*

Since Wood and McCarthy (1984) it has been regarded as an optimum when a single PCA component combines the complete between-condition variance of a single condition factor with some within-condition variance. However, the allocation of variance of a single condition factor on a single principal component combining within- and between-condition variance does not necessarily occur and the allocation of between-condition variance on more than one component has been termed 'misallocation of variance' (Wood & McCarthy, 1984). Misallocation of variance has been investigated in simulation studies on methods of PCA component rotation (e.g., Dien, 2010; Beauducel & Debener, 2003; Wood & McCarthy, 1984) and new methods of component rotation have been proposed that may reduce misallocation of variance (Beauducel, 2018; Beauducel & Leue, 2015).

It has also been proposed to perform a separate PCA for each group representing a level of the condition factor because the loading shapes in each condition can be different (Barry, De Blasio, Fogarty, Karamacoska, 2016). Although it might be reasonable to identify condition-specific loading patterns by means of separate PCAs at each level of a condition factor, the effect of this form of analysis on misallocation of variance remains unknown.

*1.3 Aims of the present paper*

The present paper therefore investigates the effects of separate PCAs at each level of a condition factor on the allocation of between-condition variance on PCA components. First, some definitions for separate PCAs at each level of a single condition factor and for a PCA of the between-condition variance of the condition factor are presented. Second, it is shown that misallocation of condition variance as it has been demonstrated and discussed since Wood and McCarthy (1984) follows necessarily from rotation of components that perfectly represent a single condition effect. Third, it is shown that different condition-specific loading shapes do not allow for an unambiguous allocation of between-condition variance on a single component representing within- and between-condition variance. Finally, it is shown that different condition-specific loading patterns are compatible with an unambiguous allocation of between-condition variance on a single component, when the between-condition differences of the loadings on each component can be accounted for by a scalar.

## 2. Definitions: PCA for within- and between-condition variance

Consider that p random variables have been observed in k levels of a condition factor, so that

$$\mathbf{x} = [\mathbf{x}_i \ \cdots \ \mathbf{x}_k], \ for \ i = 1, 2, \ldots, k. \quad (1)$$

Although the expectation of x is zero $(E[\mathbf{x}] = 0)$, the conditions imply $E[\mathbf{x}_i] \neq 0$. However, when a within-condition PCA is performed separately for the correlations or covariances at each level of the condition factor, $\mathbf{x}_i^v = \mathbf{x}_i - E[\mathbf{x}_i]$, the mean centered part of $\mathbf{x}_i$, is analyzed, since $Cov[\mathbf{x}_i, \mathbf{x}_i] = Cov[\mathbf{x}_i^v, \mathbf{x}_i^v] = \mathbf{\Sigma}_i$, so that

$$\mathbf{x}_i^v = \mathbf{A}_i^v \mathbf{c}_i^v, \ for \ i = 1, 2, \ldots k, \quad (2)$$

superscript "v" denotes the within-condition variance and where $\mathbf{A}_i^v$ is a p × p matrix of component loadings and $\mathbf{A}_i^{v'} \mathbf{A}_i^v$ contains the eigenvalues in decreasing order. The components $\mathbf{c}_i^v$ are assumed to have an expectation zero, $E[\mathbf{c}_i^v] = 0$. PCA initially yields orthogonal components $(E[\mathbf{c}_i^v \mathbf{c}_i^{v'}] = \mathbf{I})$, so that each covariance matrix of observed variables can be decomposed into

$$\mathbf{\Sigma}_i^v = \mathbf{A}_i^v \mathbf{A}_i^{v'}, \ for \ i = 1, 2, \ldots k. \quad (3)$$

Typically, components ci are divided into a subset of q wanted components wi and p – q unwanted components ui ($\mathbf{c}_i = [\mathbf{w}_i, \mathbf{u}_i]$, $\mathbf{A}_i = [\mathbf{M}_i, \mathbf{N}_i]$). Orthogonal and oblique rotations of $\mathbf{M}_i$ and $\mathbf{w}_i$ have been proposed, so that non-zero component inter-correlations are possible $(E[\mathbf{c}_i \mathbf{c}_i'] = \mathbf{Q}_i)$. The covariances of observed variables are then decomposed by

$$\mathbf{\Sigma}_i = \mathbf{M}_i \mathbf{Q}_i \mathbf{M}_i' + \mathbf{N}_i \mathbf{N}_i', \quad \text{for } i=1,2,...k. \tag{4}$$

It is possible to write the complete data comprising condition variance and within-group variance as

$$\begin{aligned}
\mathbf{x} &= \left[ \mathbf{x}_i^v + diag(\mathrm{E}[\mathbf{x}_i])\mathbf{1}_i, \quad \cdots, \quad \mathbf{x}_k^v + diag(\mathrm{E}[\mathbf{x}_k])\mathbf{1}_k \right] \\
&= \left[ \mathbf{x}_i^v \quad \cdots \quad \mathbf{x}_k^v \right] + \left[ diag(\mathrm{E}[\mathbf{x}_i])\mathbf{1}_i, \quad \cdots, \quad diag(\mathrm{E}[\mathbf{x}_k])\mathbf{1}_k \right] \\
&= \left[ \mathbf{x}_i^v \quad \cdots \quad \mathbf{x}_k^v \right] + \left[ \mathbf{x}_i^b \quad \cdots \quad \mathbf{x}_k^b \right],
\end{aligned} \tag{5}$$

where $\mathbf{1}_i$ has the dimensions of $\mathbf{x}_i$ and $\mathbf{1}_k$ has the dimensions of $\mathbf{x}_k$. The related within- and between-conditions PCAs yield

$$\mathbf{x} = \left[ \mathbf{A}_i^v \mathbf{c}_i^v, \quad \cdots, \quad \mathbf{A}_k^v \mathbf{c}_k^v \right] + \mathbf{A}^b \mathbf{c}^b. \tag{6}$$

Usually $q_v$ wanted within-condition components $\mathbf{w}_i^v$ are separated from $p - q_v$ unwanted within-condition components $\mathbf{u}_i^v$ and $q_b$ wanted between-condition components $\mathbf{w}^b$ from $p - q_b$ unwanted between-condition components $\mathbf{u}^b$. This yields

$$\mathbf{x} = \left[ \mathbf{M}_i^v \mathbf{w}_i^v + \mathbf{N}_i^v \mathbf{u}_i^v, \quad \cdots, \quad \mathbf{M}_k^v \mathbf{w}_k^v + \mathbf{N}_k^v \mathbf{u}_k^v \right] + \mathbf{M}^b \mathbf{w}^b + \mathbf{N}^b \mathbf{u}^b, \tag{7}$$

and

$$\mathbf{x}_w = \left[ \mathbf{M}_i^v \mathbf{w}_i^v, \quad \cdots, \quad \mathbf{M}_k^v \mathbf{w}_k^v \right] + \mathbf{M}^b \mathbf{w}^b, \tag{8}$$

for the wanted components.

Typically, the wanted components are rotated in order to improve the interpretation (Dien, 2010; Kayser & Tenke, 2003). If there is an additional condition factor, there can be additional groupings of PCAs for each level of the condition factor and an additional PCA across the levels of the condition factor. If the sample size is sufficiently large, it is also possible to perform a PCA for each of the combinations of condition levels and across all combinations of conditions of the two condition factors.

### 3. Misallocation of variance

*3.1 Misallocation of variance and component rotation*

When there are only a few condition factors the number of wanted within-condition components is probably larger than the number of wanted between-condition components. For example, when there is only one condition factor with two levels, PCA of the between-condition variance without subsequent component rotation will result in only one between-condition component. When $q^v > q^b = 1$ it is possible to write Equation 8 as

$$\mathbf{x}_w = \left[ \left[ \mathbf{m}_{ji}^v, \cdots, \mathbf{m}_{qi}^v \right] \begin{bmatrix} \mathbf{w}_{ji}^v \\ \vdots \\ \mathbf{w}_{qi}^v \end{bmatrix}, \cdots, \left[ \mathbf{m}_{jk}^v, \cdots, \mathbf{m}_{qk}^v \right] \begin{bmatrix} \mathbf{w}_{jk}^v \\ \vdots \\ \mathbf{w}_{qk}^v \end{bmatrix} \right] + \mathbf{m}^b \mathbf{w}^b, \tag{9}$$

where $j$ denotes the number of the respective within-condition component. For $q^b = 1$ and $\mathbf{m}_{1i}^v = \mathbf{m}^b, \cdots, \mathbf{m}_{1k}^v = \mathbf{m}^b$ Equation 9 can be written as

$$\mathbf{x}_w = \mathbf{M} \left[ \begin{bmatrix} \mathbf{w}_{ji} \\ \vdots \\ \mathbf{w}_{qi} \end{bmatrix}, \cdots, \begin{bmatrix} \mathbf{w}_{jk} \\ \vdots \\ \mathbf{w}_{qk} \end{bmatrix} \right], \tag{10}$$

with $\mathbf{M} = \left[ \mathbf{m}^b, \mathbf{m}_{2i}^v, \cdots, \mathbf{m}_{qi}^v \right]$ and $\left[ \mathbf{w}_{1i}, \cdots, \mathbf{w}_{1k} \right] = \left[ \mathbf{w}_{1i}^v + \mathbf{w}_{1i}^b, \cdots, \mathbf{w}_{1k}^v + \mathbf{w}_{1k}^b \right]$,

where $\mathbf{w}_{1i}^v, \cdots, \mathbf{w}_{1k}^v$ denotes the scores on the first wanted component ($j=1$) at each level $i$ of the condition factor, and $\mathbf{w}_{1i}^b, \cdots, \mathbf{w}_{1k}^b$ denotes the expectancy of the first wanted component on each level $i$ of the condition factor, which corresponds to the expectancy of the observed scores on condition level $i$, with $[\mathbf{w}_{1i}^b, \cdots, \mathbf{w}_{1k}^b] = [\mathrm{E}(\mathbf{w}_{1i}), \cdots, \mathrm{E}(\mathbf{w}_{1k})] = [\mathrm{E}(\mathbf{x}_i), \cdots, \mathrm{E}(\mathbf{x}_k)]$.

Equation 10 describes what is typically regarded as an optimal allocation of variance, namely, that a condition effect occurs on a single component that combines within- and between-condition variance. The simulation studies on this issue were based on a single condition effect that was introduced exclusively on a single component when the data were generated (Wood & McCarthy, 1984; Dien, 2010; Beauducel & Debener, 2003; Beauducel & Leue, 2015) and that occurred on more than one component after PCA followed by component rotation.

Component rotation means that the **M** is rotated by means of postmultiplication by a $q^v \times q^v$ transformation matrix **T** (Harman, 1976) and that the component scores are counter-rotated by means of premultiplication with $\mathbf{T}^{-1}$, so that

$$\mathbf{x}_w = \mathbf{MT}\left[\mathbf{T}^{-1}\begin{bmatrix}\mathbf{w}_{ji}\\ \vdots \\ \mathbf{w}_{qi}\end{bmatrix}, \cdots, \mathbf{T}^{-1}\begin{bmatrix}\mathbf{w}_{jk}\\ \vdots \\ \mathbf{w}_{qk}\end{bmatrix}\right], \text{ with } [\mathbf{w}_{1i},\cdots,\mathbf{w}_{1k}] = [\mathbf{w}_{1i}^v + \mathbf{w}_{1i}^b, \cdots, \mathbf{w}_{1k}^v + \mathbf{w}_{1k}^b]. \quad (11)$$

For a single condition $i$ the rotation of the infinite matrices containing the population of individual component scores $l$ can be written as

$$\mathbf{T}^{-1}\begin{bmatrix}\mathbf{w}_{ji}\\ \vdots \\ \mathbf{w}_{qi}\end{bmatrix} = \mathbf{T}^*\begin{bmatrix}\mathbf{w}_{jil},\cdots \\ \vdots \ddots \\ \mathbf{w}_{qil},\cdots\end{bmatrix} = \begin{bmatrix}(\mathbf{t}^*\mathbf{w})_{jil},\cdots \\ \vdots \ddots \\ (\mathbf{t}^*\mathbf{w})_{qil},\cdots\end{bmatrix}, \quad (12)$$

for $l = 1,\ldots,\infty$, with $[\mathbf{w}_{1il}, \cdots] = [\mathbf{w}_{1il}^v + \mathrm{E}(\mathbf{w}_{1i}), \cdots]$ and $\mathbf{T}^{-1} = \mathbf{T}^*$.

Theorem 1 describes that a non-zero expectation that is initially only on the first component leads to a non-zero expectation on others than the first component after component rotation.

**Theorem 1.** if $\mathrm{E}(\mathbf{w}_{ji}) = \begin{cases}\mathrm{E}(\mathbf{w}_{ji}) \neq 0 \text{ for } j = 1 \\ \mathrm{E}(\mathbf{w}_{ji}) = 0 \text{ for } j = 2,\ldots,q\end{cases}$ and $\mathbf{t}_{jh}^* \neq 0$, for $j = 1,\ldots,q, h = 1,\ldots,q$

then $\mathrm{E}(\mathbf{t}^*\mathbf{w})_{ji} \neq 0$, for $j > 1,\ldots,q$.

*Proof.* A single element for condition $i$ of the matrix resulting from Equation 12 is given by

$$(\mathbf{t}^*\mathbf{w})_{jil} = \sum_{h=1}^{q}\mathbf{t}_{jh}^*\mathbf{w}_{hil}, \text{ with } \mathbf{w}_{1il} = \mathbf{w}_{1il}^v + \mathrm{E}(\mathbf{w}_{1i}). \quad (13)$$

Equation 13 can be written as

$$(\mathbf{t}^*\mathbf{w})_{jil} = \mathbf{t}_{j1}^*(\mathbf{w}_{1il}^v + \mathrm{E}(\mathbf{w}_{1i})) + ,\ldots, + \mathbf{t}_{jq}^*\mathbf{w}_{qil}. \quad (14)$$

Equation 14 implies that the expectation for the population of scores even for $j > 1$ is $\mathrm{E}(\mathbf{t}^*\mathbf{w})_{ji} = \mathbf{t}_{j1}^*\mathrm{E}(\mathbf{w}_{1i})$.

This completes the proof. □

Theorem 1 implies that a condition effect that occurs only on the first component before rotation, also occurs on other components after rotation. Thus, Theorem 1 shows that misallocation of variance as it has typically been investigated in simulation studies since Wood and McCarthy (1984) is a necessary consequence of any rotation of an initial set of components combining unambiguously within- and between-condition effects. Therefore, the attempts to reduce misallocation of variance are attempts to recover the initial combination of within- and between-condition components (Dien, 2010; Beauducel & Leue, 2015; Beauducel, 2018) so that the matrix **T**, transforming the original components to the given components becomes **I**. This implies $\mathbf{T}^* = \mathbf{I}$ and $\mathbf{t}_{jh}^* = 0$, for $j \neq h$ so that Theorem 1 does not hold. Eliminating variance misallocation by means of component rotation precludes that there exists a PCA solution for the data at hand where each between-condition effect can be allocated on a separate single component. This is, however, not necessarily the case for any data set.

*3.2 Misallocation of variance in combined within- and between-condition components*

Theorem 1 describes misallocation of variance as it can occur when PCA is performed for the total sample, i.e., across the levels of a between-condition factor. When separate within-condition components $\mathbf{c}_i^v,\ldots,\mathbf{c}_k^v$ are computed, the within-condition components $\mathbf{c}_i^v,\ldots,\mathbf{c}_k^v$ are completely unrelated to $\mathbf{c}^b$ so that within- and between-condition variance is completely disentangled. This yields the question under which constraints within-and between-condition components can be combined into a single component representing within- and between-condition variance unambiguously. Theorem 2 describes a constraint for the component loadings that implies $\mathbf{c} = [\mathbf{c}_i^v + \mathbf{c}_i^b, \cdots, \mathbf{c}_k^v + \mathbf{c}_k^b]$, i.e., that each component in $\mathbf{c}$ can be decomposed into a separate within- and between-condition component. This implies that no misallocation of variance occurs because each between-condition component is uniquely combined with another within-condition component.

**Theorem 2.** *If* $\mathbf{A}_i^v = \mathbf{A}^b, \cdots, \mathbf{A}_k^v = \mathbf{A}^b$ *then* $\mathbf{c} = [\mathbf{c}_i^v, \cdots, \mathbf{c}_k^v] + [\mathbf{c}_i^b, \cdots, \mathbf{c}_k^b]$.

*Proof.* Since $\mathbf{c}^b = [\mathbf{c}_i^b, \ldots, \mathbf{c}_k^b]$ Equation 6 can be written as

$$\mathbf{x} = [\mathbf{A}_i^v \mathbf{c}_i^v, \cdots, \mathbf{A}_k^v \mathbf{c}_k^v] + \mathbf{A}^b [\mathbf{c}_i^b, \cdots, \mathbf{c}_k^b]. \tag{15}$$

Inserting $\mathbf{A}^b$ for $\mathbf{A}_i^v, \cdots, \mathbf{A}_k^v$ into Equation 15 yields

$$\begin{aligned}\mathbf{x} &= \mathbf{A}^b [\mathbf{c}_i^v, \cdots, \mathbf{c}_k^v] + \mathbf{A}^b [\mathbf{c}_i^b, \cdots, \mathbf{c}_k^b] \\ &= \mathbf{A}^b [\mathbf{c}_i^v + \mathbf{c}_i^b, \cdots, \mathbf{c}_k^v + \mathbf{c}_k^b] = \mathbf{A}^b \mathbf{c}.\end{aligned} \tag{16}$$

This completes the proof. □

Thus, when the within-condition loading matrices at each condition level are identical to the between-condition loading matrix, this implies a component model where all components combine their respective within- and between-condition variance. Theorem 2 implies that no misallocation of variance occurs when each condition-specific loading pattern is identical to the between-condition loading pattern. When Theorem 2 holds, it would be possible to find a solution without variance misallocation by means of component rotation.

Writing loading vectors $\mathbf{A}_i^v = [\mathbf{a}_{si}^v, \cdots, \mathbf{a}_{pi}^v], \mathbf{A}^b = [\mathbf{a}_s^b, \cdots, \mathbf{a}_p^b]$ and component score vectors $\mathbf{c}_i^v = \begin{bmatrix} \mathbf{c}_{si}^v \\ \vdots \\ \mathbf{c}_{pi}^v \end{bmatrix}, \mathbf{c}_i^b = \begin{bmatrix} \mathbf{c}_{si}^b \\ \vdots \\ \mathbf{c}_{pi}^b \end{bmatrix}$ for the $s$ to $p$ components in Equation 16 yields

$$\mathbf{x} = \left[ [\mathbf{a}_{si}^v, \cdots, \mathbf{a}_{pi}^v] \begin{bmatrix} \mathbf{c}_{si}^v \\ \vdots \\ \mathbf{c}_{pi}^v \end{bmatrix}, \cdots, [\mathbf{a}_{sk}^v, \cdots, \mathbf{a}_{pk}^v] \begin{bmatrix} \mathbf{c}_{sk}^v \\ \vdots \\ \mathbf{c}_{pk}^v \end{bmatrix} \right] + [\mathbf{a}_s^b, \cdots, \mathbf{a}_p^b] \left[ \begin{bmatrix} \mathbf{c}_{si}^b \\ \vdots \\ \mathbf{c}_{pi}^b \end{bmatrix}, \cdots, \begin{bmatrix} \mathbf{c}_{sk}^b \\ \vdots \\ \mathbf{c}_{pk}^b \end{bmatrix} \right]. \tag{17}$$

Note that the scores $\mathbf{c}_{si}^b$ are equal for each between-condition component $s$ at each condition-level $i$. For convenience, the raw data reproduced from the first component are considered. This yields

$$\mathbf{x}_1^* = [\mathbf{a}_{1i}^v \mathbf{c}_{1i}^v, \cdots, \mathbf{a}_{1k}^v \mathbf{c}_{1k}^v] + \mathbf{a}_1^b [\mathbf{c}_{1i}^b, \cdots, \mathbf{c}_{1k}^b]. \tag{18}$$

It follows from $\mathbf{a}_{1i}^v = \mathbf{a}_1^b, \cdots, \mathbf{a}_{1k}^v = \mathbf{a}_1^b$ that $\mathbf{c}_1 = [\mathbf{c}_{1i}^v + \mathbf{c}_{1i}^b, \cdots, \mathbf{c}_{1k}^v + \mathbf{c}_{1k}^b]$ and that $\mathbf{x}_1^* = \mathbf{a}_1^b \mathbf{c}_1$. Thus, it is possible that only a subset of the within-condition loading matrices and between-condition loading matrices is identical and that this subset of components combines within- and between-condition variance. When there is only one between-condition component, i.e., $q^v = p > q^b = 1$, Equation 17 can be written as

$$\mathbf{x} = \left[ [\mathbf{a}_{si}^v, \cdots, \mathbf{a}_{pi}^v] \begin{bmatrix} \mathbf{c}_{si}^v \\ \vdots \\ \mathbf{c}_{pi}^v \end{bmatrix}, \cdots, [\mathbf{a}_{sk}^v, \cdots, \mathbf{a}_{pk}^v] \begin{bmatrix} \mathbf{c}_{sk}^v \\ \vdots \\ \mathbf{c}_{pk}^v \end{bmatrix} \right] + \mathbf{a}_1^b [\mathbf{c}_{1i}^b, \cdots, \mathbf{c}_{1k}^b]. \tag{19}$$

Theorem 3 describes constraints for the loadings that are compatible with a model combining a single between-condition component with the first within-condition component.

**Theorem 3.** *If* $q^v = p > q^b = 1$, *and* $\mathbf{a}_{1i}^v = \mathbf{a}_1^b, \cdots, \mathbf{a}_{1k}^v = \mathbf{a}_1^b$ *and* $\mathbf{a}_{si}^v \neq \mathbf{a}_1^b, \cdots, \mathbf{a}_{sk}^v \neq \mathbf{a}_1^b$ *then* $\mathbf{c}_1 = [\mathbf{c}_{1i}^v + \mathbf{c}_{1i}^b, \cdots, \mathbf{c}_{1k}^v + \mathbf{c}_{1k}^b]$ *and* $\mathbf{c}_s \neq [\mathbf{c}_{si}^v + \mathbf{c}_{si}^b, \cdots, \mathbf{c}_{sk}^v + \mathbf{c}_{sk}^b]$, *for* $s = 2, \ldots, p$.

*Proof.* For $\mathbf{a}_{1i}^v = \mathbf{a}_1^b, \cdots, \mathbf{a}_{1k}^v = \mathbf{a}_1^b$ Equation 18 can be written as

$$\mathbf{x} = \left[ [\mathbf{a}_1^b, \mathbf{a}_{si}^v, \cdots, \mathbf{a}_{pi}^v] \begin{bmatrix} \mathbf{c}_{1i}^v \\ \mathbf{c}_{si}^v \\ \vdots \\ \mathbf{c}_{pi}^v \end{bmatrix}, \quad \cdots, [\mathbf{a}_1^b, \mathbf{a}_{sk}^v, \cdots, \mathbf{a}_{pk}^v] \begin{bmatrix} \mathbf{c}_{1k}^v \\ \mathbf{c}_{sk}^v \\ \vdots \\ \mathbf{c}_{pk}^v \end{bmatrix} \right] + \mathbf{a}_1^b \left[ \mathbf{c}_{1i}^b, \quad \cdots, \mathbf{c}_{1k}^b \right]$$

$$= \left[ [\mathbf{a}_{si}^v, \cdots, \mathbf{a}_{pi}^v] \begin{bmatrix} \mathbf{c}_{si}^v \\ \vdots \\ \mathbf{c}_{pi}^v \end{bmatrix}, \quad \cdots, [\mathbf{a}_{sk}^v, \cdots, \mathbf{a}_{pk}^v] \begin{bmatrix} \mathbf{c}_{sk}^v \\ \vdots \\ \mathbf{c}_{pk}^v \end{bmatrix} \right] + \mathbf{a}_1^b \mathbf{c}_1, \qquad (20)$$

with $\mathbf{c}_1 = \left[ \mathbf{c}_{1i}^v + \mathbf{c}_{1i}^b, \quad \cdots, \mathbf{c}_{1k}^v + \mathbf{c}_{1k}^b \right]$, for $s = 2, ..., p$.

This completes the proof. □

The identity of the loading patterns of the first unrotated within- and between-condition components is a necessary constraint for the allocation of the between- and within-condition variance on a common component. Theorem 4 describes a somewhat relaxed constraint that is based on an identical shape of the loadings of the first within- and between-condition components but allows for a different scale.

**Theorem 4.** If $q^v = p > q^b = 1$, and $\theta_i \mathbf{a}_{1i}^v = \mathbf{a}_1^b, \cdots, \theta_k \mathbf{a}_{1k}^v = \mathbf{a}_1^b$ and $\theta_i \mathbf{a}_{si}^v \neq \mathbf{a}_1^b, \cdots, \theta_k \mathbf{a}_{sk}^v \neq \mathbf{a}_1^b$,

then $\mathbf{c}_1 = \left[ \theta_i \mathbf{c}_{1i}^v + \mathbf{c}_{1i}^b, \quad \cdots, \theta_k \mathbf{c}_{1k}^v + \mathbf{c}_{1k}^b \right]$ and $\mathbf{c}_s \neq \left[ \theta_i \mathbf{c}_{si}^v + \mathbf{c}_{si}^b, \quad \cdots, \theta_k \mathbf{c}_{sk}^v + \mathbf{c}_{sk}^b \right]$, for $s = 2, ..., p$ and $\theta_i > 0, \cdots, \theta_k > 0$.

*Proof.* For $\theta_i \mathbf{a}_{1i}^v = \mathbf{a}_1^b, \cdots, \theta_k \mathbf{a}_{1k}^v = \mathbf{a}_1^b$ Equation 18 can be written as

$$\mathbf{x} = \left[ [\theta_i \mathbf{a}_1^b, \mathbf{a}_{si}^v, \cdots, \mathbf{a}_{pi}^v] \begin{bmatrix} \mathbf{c}_{1i}^v \\ \mathbf{c}_{si}^v \\ \vdots \\ \mathbf{c}_{pi}^v \end{bmatrix}, \quad \cdots, [\theta_k \mathbf{a}_1^b, \mathbf{a}_{sk}^v, \cdots, \mathbf{a}_{pk}^v] \begin{bmatrix} \mathbf{c}_{1k}^v \\ \mathbf{c}_{sk}^v \\ \vdots \\ \mathbf{c}_{pk}^v \end{bmatrix} \right] + \mathbf{a}_1^b \left[ \mathbf{c}_{1i}^b, \quad \cdots, \mathbf{c}_{1k}^b \right]$$

$$= \left[ [\mathbf{a}_{si}^v, \cdots, \mathbf{a}_{pi}^v] \begin{bmatrix} \mathbf{c}_{si}^v \\ \vdots \\ \mathbf{c}_{pi}^v \end{bmatrix}, \quad \cdots, [\mathbf{a}_{sk}^v, \cdots, \mathbf{a}_{pk}^v] \begin{bmatrix} \mathbf{c}_{sk}^v \\ \vdots \\ \mathbf{c}_{pk}^v \end{bmatrix} \right] + \mathbf{a}_1^b \mathbf{c}_1, \qquad (21)$$

with $\mathbf{c}_1 = \left[ \theta_i \mathbf{c}_{1i}^v + \mathbf{c}_{1i}^b, \quad \cdots, \theta_k \mathbf{c}_{1k}^v + \mathbf{c}_{1k}^b \right]$, for $s = 2, ...., p$.

This completes the proof. □

Theorem 4 shows that condition-specific loading patterns that have the same shape, but a different scale are compatible with a model where a single between-condition component is unambiguously allocated on a single within-condition component.

## 4. Discussion

According to Wood and McCarthy (1984) misallocation of variance occurs when a single between-condition effect that can in principle be allocated on a single PCA component is allocated on more than one component in a given PCA solution. The present study describes constraints that are to be imposed on the component loading matrices in order to avoid misallocation of variance. The following conclusions can be drawn: When a single between-condition effect is allocated on a single component of an initial PCA solution, any rotation of these initial components will result in a misallocation of variance (Theorem 1). This is an algebraic demonstration of what has been discussed elsewhere (Dien, 2010; Beauducel & Leue, 2015; Beauducel, 2018), namely that, at the level of combined within- and between-condition components, the misallocation of variance is directly related to component rotation. However, component rotation can only result in an optimal allocation of between-condition variance when such a rotational solution exists for a given data set.

Since it has been proposed to perform separate PCAs at each level of a condition factor (Barry et al., 2016), the consequences of this procedure for misallocation of variance were explored. When a PCA is calculated at each level of a condition factor and when a PCA is calculated for a single between-condition factor, an unambiguous allocation of the between-condition variance on a single component combining within- and between-condition variance is possible when the within-condition component loadings have the same shape, even when their scale is different (Theorem 3 and 4). Thus, only when the constraints given Theorem 3 and 4 hold for a given data set, it would be possible to find the solution with optimal allocation of between-condition variance by means of component rotation.

Theorem 3 and 4 also imply that separate PCAs at each level of a condition-factor are not necessarily a way to avoid or eliminate misallocation of variance. When different loading shapes occur at each level of a condition factor in separate PCAs, this indicates that misallocation of variance would occur when the separate components are combined into within- and between variance components. In contrast, when the loading shape is similar in the different PCAs with larger or smaller loadings at each level of the condition factor, the components can be combined into within- and between-components without misallocation of variance.

Finally, it follows from Theorem 4 that perfect congruence coefficients (Tucker, 1951; Wrigley, & Neuhaus, 1955) of the loadings of respective components at different levels of the condition factor are not a necessary condition for optimal variance allocation because congruence coefficients also refer to the similarity of the loading magnitude. For optimal variance allocation, a perfect Pearson correlation of the loadings of the respective components at different levels of the condition factor would be sufficient.